\documentclass[conference]{IEEEtran}
\usepackage{amsmath}

\usepackage{xspace}
\usepackage{hyperref}
\usepackage[hyphenbreaks]{breakurl}

\usepackage{subfigure}
\usepackage{psfrag}
\usepackage{epsfig}
\usepackage{tabularx}
\usepackage{graphicx}
\usepackage{multirow}
\usepackage{algorithm,algpseudocode}
\usepackage[english]{babel}
\usepackage{array}
\usepackage{algpseudocode}
\usepackage{url}

\usepackage{psfrag}
\usepackage{pstricks,pst-node}
\usepackage{pstricks-add}
\usepackage{bm}
\usepackage{amssymb}
\usepackage{tikz}
\usepackage{enumitem}
\usepackage{pst-blur}

\usepackage{listings}
\usepackage{color}

\definecolor{dkgreen}{rgb}{0,0.6,0}
\definecolor{gray}{rgb}{0.5,0.5,0.5}
\definecolor{mauve}{rgb}{0.58,0,0.82}

\usepackage{listings} \newcommand{\pgftextcircled}[1]{
    \setbox0=\hbox{#1}%
    \dimen0\wd0%
    \divide\dimen0 by 2%
    \begin{tikzpicture}[baseline=(a.base)]%
        \useasboundingbox (-\the\dimen0,0pt) rectangle (\the\dimen0,1pt);
        \node[circle,draw,outer sep=0pt,inner sep=0.1ex] (a) {#1};
    \end{tikzpicture}
}

\graphicspath{{figs/}{experiment/result/figure/}}

\usepackage{listings} \newcommand{\todo}[1]{%
    \ifthenelse{\boolean{clean}}{}{
        \noindent\color{blue}[{\small{$\star$~ToDo(#1)}\/}]%
        \color{black}
    }
}

\newcommand{\secref}[1]{Section~\ref{#1}}

\newcommand{\figref}[1]{Fig.~\ref{#1}}
\newcommand{\tabref}[1]{Tab.~\ref{#1}}
\newcommand{\eqnref}[1]{Eqn.~(\ref{#1})}

\newcommand{\V}{\ensuremath{\mathcal{V}}\xspace}

\newcommand{\GA}{\gamma}
\newcommand{\GAH}{\hat{\gamma}}
\newcommand{\KA}{\kappa}
\newcommand{\PII}{\pi}
\newcommand{\ALP}{\alpha}
\newcommand{\ALPB}{\bar{\alpha}}
\newcommand{\ALPBH}{\hat{\bar{\alpha}}}
\newcommand{\BE}{\beta}
\newcommand{\BEB}{\bar{\beta}}
\newcommand{\SIG}{\sigma}
\newcommand{\SIB}{\bar{\sigma}}

\newcommand{\inv}{^{-1}}
\newcommand{\VCC}[1][noarg]{\ifthenelse{\equal{#1}{noarg}}{\ensuremath{\V}}{\ensuremath{\V_{#1}}}\xspace}
\newcommand{\VCCi}[1][noarg]{\ifthenelse{\equal{#1}{noarg}}{\ensuremath{\V\inv}}{\ensuremath{\V_{#1}\inv}}\xspace}
\newcommand{\uVCC}[1][noarg]{\ifthenelse{\equal{#1}{noarg}}{\ensuremath{\V^u}}{\ensuremath{\V_{#1}^u}}\xspace}
\newcommand{\lVCC}[1][noarg]{\ifthenelse{\equal{#1}{noarg}}{\ensuremath{\V^l}}{\ensuremath{\V_{#1}^l}}\xspace}
\newcommand{\uVCCi}[1][noarg]{\ifthenelse{\equal{#1}{noarg}}{\ensuremath{\V^{u\inv}}}{\ensuremath{\V_{#1}^{u\inv}}}\xspace}
\newcommand{\lVCCi}[1][noarg]{\ifthenelse{\equal{#1}{noarg}}{\ensuremath{\V^{l\inv}}}{\ensuremath{\V_{#1}^{l\inv}}}\xspace}

\newcommand{\G}[1][noarg]{\ifthenelse{\equal{#1}{noarg}}{\ensuremath{\GA}}{\ensuremath{\GA_{#1}}}\xspace}
\newcommand{\uG}[1][noarg]{\ifthenelse{\equal{#1}{noarg}}{\ensuremath{\GA^u}}{\ensuremath{\GA_{#1}^u}}\xspace}
\newcommand{\lG}[1][noarg]{\ifthenelse{\equal{#1}{noarg}}{\ensuremath{\GA^l}}{\ensuremath{\GA_{#1}^l}}\xspace}
\newcommand{\Gi}[1][noarg]{\ifthenelse{\equal{#1}{noarg}}{\ensuremath{\GA\inv}}{\ensuremath{\GA_{#1}\inv}}\xspace}
\newcommand{\uGi}[1][noarg]{\ifthenelse{\equal{#1}{noarg}}{\ensuremath{\GA^{u\inv}}}{\ensuremath{\GA_{#1}^{u\inv}}}\xspace}
\newcommand{\lGi}[1][noarg]{\ifthenelse{\equal{#1}{noarg}}{\ensuremath{\GA^{l\inv}}}{\ensuremath{\GA_{#1}^{l\inv}}}\xspace}
\newcommand{\uGh}[1][noarg]{\ifthenelse{\equal{#1}{noarg}}{\ensuremath{\GAH^u}}{\ensuremath{\GAH_{#1}^u}}\xspace}

\newcommand{\K}[1][noarg]{\ifthenelse{\equal{#1}{noarg}}{\ensuremath{\KA}}{\ensuremath{\KA_{#1}}}\xspace}
\newcommand{\uK}[1][noarg]{\ifthenelse{\equal{#1}{noarg}}{\ensuremath{\KA^u}}{\ensuremath{\KA_{#1}^u}}\xspace}
\newcommand{\lK}[1][noarg]{\ifthenelse{\equal{#1}{noarg}}{\ensuremath{\KA^l}}{\ensuremath{\KA_{#1}^l}}\xspace}
\newcommand{\uKi}[1][noarg]{\ifthenelse{\equal{#1}{noarg}}{\ensuremath{\KA^{u\inv}}}{\ensuremath{\KA_{#1}^{u\inv}}}\xspace}
\newcommand{\lKi}[1][noarg]{\ifthenelse{\equal{#1}{noarg}}{\ensuremath{\KA^{l\inv}}}{\ensuremath{\KA_{#1}^{l\inv}}}\xspace}

\newcommand{\PI}[1][noarg]{\ifthenelse{\equal{#1}{noarg}}{\ensuremath{\PII}}{\ensuremath{\PII_{#1}}}\xspace}
\newcommand{\uP}[1][noarg]{\ifthenelse{\equal{#1}{noarg}}{\ensuremath{\PII^u}}{\ensuremath{\PII_{#1}^u}}\xspace}
\newcommand{\lP}[1][noarg]{\ifthenelse{\equal{#1}{noarg}}{\ensuremath{\PII^l}}{\ensuremath{\PII_{#1}^l}}\xspace}
\newcommand{\uPi}[1][noarg]{\ifthenelse{\equal{#1}{noarg}}{\ensuremath{\PII^{u\inv}}}{\ensuremath{\PII_{#1}^{u\inv}}}\xspace}
\newcommand{\lPi}[1][noarg]{\ifthenelse{\equal{#1}{noarg}}{\ensuremath{\PII^{l\inv}}}{\ensuremath{\PII_{#1}^{l\inv}}}\xspace}

\newcommand{\AL}[1][noarg]{\ifthenelse{\equal{#1}{noarg}}{\ensuremath{\ALP}}{\ensuremath{\ALP_{#1}}}\xspace}
\newcommand{\Ab}[1][noarg]{\ifthenelse{\equal{#1}{noarg}}{\ensuremath{\ALPB}}{\ensuremath{\ALPB_{#1}}}\xspace}
\newcommand{\uA}[1][noarg]{\ifthenelse{\equal{#1}{noarg}}{\ensuremath{\ALP^u}}{\ensuremath{\ALP_{#1}^u}}\xspace}
\newcommand{\lA}[1][noarg]{\ifthenelse{\equal{#1}{noarg}}{\ensuremath{\ALP^l}}{\ensuremath{\ALP_{#1}^l}}\xspace}
\newcommand{\uAi}[1][noarg]{\ifthenelse{\equal{#1}{noarg}}{\ensuremath{\ALP^{u\inv}}}{\ensuremath{\ALP_{#1}^{u\inv}}}\xspace}
\newcommand{\lAi}[1][noarg]{\ifthenelse{\equal{#1}{noarg}}{\ensuremath{\ALP^{l\inv}}}{\ensuremath{\ALP_{#1}^{l\inv}}}\xspace}
\newcommand{\uAb}[1][noarg]{\ifthenelse{\equal{#1}{noarg}}{\ensuremath{\ALPB^u}}{\ensuremath{\ALPB_{#1}^u}}\xspace}
\newcommand{\lAb}[1][noarg]{\ifthenelse{\equal{#1}{noarg}}{\ensuremath{\ALPB^l}}{\ensuremath{\ALPB_{#1}^l}}\xspace}
\newcommand{\uAbi}[1][noarg]{\ifthenelse{\equal{#1}{noarg}}{\ensuremath{\ALPB^{u\inv}}}{\ensuremath{\ALPB_{#1}^{u\inv}}}\xspace}
\newcommand{\lAbi}[1][noarg]{\ifthenelse{\equal{#1}{noarg}}{\ensuremath{\ALPB^{l\inv}}}{\ensuremath{\ALPB_{#1}^{l\inv}}}\xspace}
\newcommand{\uAbh}[1][noarg]{\ifthenelse{\equal{#1}{noarg}}{\ensuremath{\ALPBH^u}}{\ensuremath{\ALPBH_{#1}^u}}\xspace}

\newcommand{\B}[1][noarg]{\ifthenelse{\equal{#1}{noarg}}{\ensuremath{\BE}}{\ensuremath{\BE_{#1}}}\xspace}
\newcommand{\Bb}[1][noarg]{\ifthenelse{\equal{#1}{noarg}}{\ensuremath{\BEB}}{\ensuremath{\BEB_{#1}}}\xspace}
\newcommand{\uB}[1][noarg]{\ifthenelse{\equal{#1}{noarg}}{\ensuremath{\BE^u}}{\ensuremath{\BE_{#1}^u}}\xspace}
\newcommand{\lB}[1][noarg]{\ifthenelse{\equal{#1}{noarg}}{\ensuremath{\BE^l}}{\ensuremath{\BE_{#1}^l}}\xspace}
\newcommand{\uBi}[1][noarg]{\ifthenelse{\equal{#1}{noarg}}{\ensuremath{\BE^{u\inv}}}{\ensuremath{\BE_{#1}^{u\inv}}}\xspace}
\newcommand{\lBi}[1][noarg]{\ifthenelse{\equal{#1}{noarg}}{\ensuremath{\BE^{l\inv}}}{\ensuremath{\BE_{#1}^{l\inv}}}\xspace}
\newcommand{\uBb}[1][noarg]{\ifthenelse{\equal{#1}{noarg}}{\ensuremath{\BEB^u}}{\ensuremath{\BEB_{#1}^u}}\xspace}
\newcommand{\lBb}[1][noarg]{\ifthenelse{\equal{#1}{noarg}}{\ensuremath{\BEB^l}}{\ensuremath{\BEB_{#1}^l}}\xspace}
\newcommand{\uBbi}[1][noarg]{\ifthenelse{\equal{#1}{noarg}}{\ensuremath{\BEB^{u\inv}}}{\ensuremath{\BEB_{#1}^{u\inv}}}\xspace}
\newcommand{\lBbi}[1][noarg]{\ifthenelse{\equal{#1}{noarg}}{\ensuremath{\BEB^{l\inv}}}{\ensuremath{\BEB_{#1}^{l\inv}}}\xspace}

\newcommand{\Sb}[1][noarg]{\ifthenelse{\equal{#1}{noarg}}{\ensuremath{\SIB}}{\ensuremath{\SIB_{#1}}}\xspace}
\newcommand{\uS}[1][noarg]{\ifthenelse{\equal{#1}{noarg}}{\ensuremath{\SIG^u}}{\ensuremath{\SIG_{#1}^u}}\xspace}
\newcommand{\lS}[1][noarg]{\ifthenelse{\equal{#1}{noarg}}{\ensuremath{\SIG^l}}{\ensuremath{\SIG_{#1}^l}}\xspace}
\newcommand{\uSb}[1][noarg]{\ifthenelse{\equal{#1}{noarg}}{\ensuremath{\SIB^u}}{\ensuremath{\SIB_{#1}^u}}\xspace}
\newcommand{\lSb}[1][noarg]{\ifthenelse{\equal{#1}{noarg}}{\ensuremath{\SIB^l}}{\ensuremath{\SIB_{#1}^l}}\xspace}

\newcommand{\task}[1][noarg]{\ifthenelse{\equal{#1}{noarg}}{\ensuremath{\mathrm{T}}}{\ensuremath{\mathrm{T{#1}}}}\xspace}

\newcommand{\PE}[1][noarg]{\ifthenelse{\equal{#1}{noarg}}{\ensuremath{\mathrm{PE}}}{\ensuremath{\mathrm{PE{#1}}}}\xspace}

\newboolean{clean}\setboolean{clean}{false}

\newcommand{\storyline}[1]{%
    \ifthenelse{\boolean{clean}}{}{
        \noindent\color{blue}{$\bigstar$ #1\/}%
        \color{black}
    }
}

\newcommand{\p}[1]{%
    \ifthenelse{\boolean{clean}}{}{
        \storyline{#1}
    }
}

\newcommand{\ask}[1]{%
    \ifthenelse{\boolean{clean}}{}{
        \color{blue}[{\small\it\textsc{$\bigstar$ (?):} #1 \/}]%
        \color{black}
    }
}%

\newcommand{\trans}[1]{%
    \ifthenelse{\boolean{clean}}{}{
        \color{green}#1%
        \color{black}
    }
}%

\newcommand{\pic}[1]{%
    \ifthenelse{\boolean{clean}}{}{
        \begin{center}%
        \color{red}[{\small\it\textsc{ Picture:} #1\/}]%
        \color{black}
        \end{center}
    }
}%

\newcommand{\ideas}{%
    \ifthenelse{\boolean{clean}}{}{
        \begin{center}
            \vspace{5mm}%
            \color{red}\texttt{\footnotesize --------------------} {\it\small ideas} \texttt{\footnotesize --------------------}
            \vspace{2mm}%
        \end{center}%
    }
}%

\newcommand{\testtext}{%
    \ifthenelse{\boolean{clean}}{}{
    \color{magenta}
    The brown {\bf fox} jumps over the lazy \emph{dog}. The quick
    brown fox jumps over the lazy dog. The quick brown fox jumps over
    the lazy dog. The quick brown fox jumps over the lazy dog. The
    quick brown fox jumps over the lazy dog.
    \color{black}
    }
}

\newcommand{\w}[1]{
    \ifthenelse{\boolean{clean}}{}{
        \underline{#1}%
    }
}%

\DeclareMathSizes{9}{8}{1}{1}
\begin{document}
%
\IEEEoverridecommandlockouts

\title{PhoneBit: Efficient GPU-Accelerated Binary Neural Network Inference Engine for Mobile Phones\thanks{This paper has been submitted to Design, Automation and Test in Europe(DATE) on Sept-15th-2019, and accepted on Nov-7th-2019. Email: cheng83@mail.sysu.edu.cn}
\vskip -0.3cm}

\author{
	\IEEEauthorblockN{Gang Chen\IEEEauthorrefmark{1}, Shengyu He\IEEEauthorrefmark{2}\IEEEauthorrefmark{3}, Haitao Meng\IEEEauthorrefmark{2}, Kai Huang\IEEEauthorrefmark{1}} \IEEEauthorblockA{\IEEEauthorrefmark{1}Sun Yat-sen University, China}\IEEEauthorblockA{\IEEEauthorrefmark{2}Notheastern University, China}\IEEEauthorblockA{\IEEEauthorrefmark{3} Peng Cheng Laboratory, China}
}

\maketitle

\def \lo {low-criticality }
\def \hi {high-criticality }
\def \FMC {FMC }
\def \FMCN {FMC-MST }
\def \prealg {FG}
\def \micr {mission-critical }
\def \noncr {non-critical }
\newcommand{\sups}[1]{\sup\{#1\}}
\newcommand{\supst}[1]{\overline{\sup}\{#1\}}
\begin{abstract}
Over the last years, a great success of deep neural networks (DNNs) has been witnessed in computer vision and other fields. However, performance and power constraints make it still challenging to deploy DNNs on mobile devices due to their high computational complexity. Binary neural networks (BNNs) have been demonstrated as a promising solution to achieve this goal by using bit-wise operations to replace most arithmetic operations. Currently, existing GPU-accelerated implementations of BNNs are \textit{only} tailored for desktop platforms. Due to architecture differences, mere porting of such implementations to \textit{mobile devices} yields suboptimal performance or is impossible in some cases. In this paper, we propose PhoneBit, a GPU-accelerated BNN inference engine for Android-based mobile devices that fully exploits the computing power of BNNs on mobile GPUs. PhoneBit provides a set of operator-level optimizations including locality-friendly data layout, bit packing with vectorization and layers integration for efficient binary convolution. We also provide a detailed implementation and parallelization optimization for PhoneBit to optimally utilize the memory bandwidth and computing power of mobile GPUs. We evaluate PhoneBit with AlexNet, YOLOv2 Tiny and VGG16 with their binary version. Our experiment results show that PhoneBit can achieve significant speedup and energy efficiency compared with state-of-the-art frameworks for mobile devices.  

\end{abstract}
\IEEEpeerreviewmaketitle
\let\olditem\item
\renewcommand{\item}{\setlength{\itemsep}{0pt}\setlength{\parsep}{0pt}\setlength{\parskip}{0pt}\olditem}

\setlength{\intextsep}{1pt}

\section{Introduction}
\label{sec:int}
In the past years, deep neural networks~(DNNs) have brought great opportunities and revolutions for many intelligence applications such as automated vehicles and natural language processing. A great success of DNN has been achieved in boosting the performance of the classification accuracy. Being ubiquitous, there is an increasing interest in applying DNNs to mobile environments. DNNs cannot only enhance the performance of mobile applications, but also pave the way toward more intelligent uses of mobile devices~\cite{Ota2017Deep}. Therefore, as an important impetus towards mobile intelligence, many recent developments in deep learning are tightly connected to tasks meant for mobile devices.

Despite the fact that DNNs are highly useful when deployed on high-end devices, deploying DNNs on mobile devices is still a challenging task because DNNs require too much memory and computing power which significantly exceeds the resource capabilities of current mobile devices and will drain out the battery soon. In addition, DNN continues to get deeper and larger. This trend makes the deployment of DNNs on mobile devices even more challenging. As a consequence, there is an increasing research interest in reducing the computational and memory requirements of DNNs~\cite{Sze2017Efficient}. 

%
%
%

Binary neural networks (BNNs) proposed in~\cite{Courbariaux2016Binarized} have been demonstrated as a promising solution for mobile computing due to their significantly reduced model size and the complexity of arithmetic operations. Using bit-wise operations to replace floating-point operations, BNN can achieve a significant model compression as well as speedup on accelerating the inference at the cost of small accuracy loss. Despite these results are highly promising, GPU-accelerated BNN inference engine and its highly optimized implementation are still not available yet for mobile devices. While there exists a number of designed deep learning frameworks~\cite{Latifi2016CNNdroid,tflite,caffe2,PaddlePaddle} for mobile devices, most of them are however designed for accelerating general CNN models with full or half precision~\cite{ji2019hg}.
In the original paper of BNN~\cite{Courbariaux2016Binarized,rastegariECCV16}, \textit{only} a proof-of-concept implementation has been provided to show the performance of BNN. In the implementation, binary weights and activations in BNN are still represented by floating-point values for proof-of-concept purposes. Recently, Pedersoli et al.~\cite{Espresso} designed an BNN inference library which was written in CUDA and tailored \textit{only} for desktop platforms. However, because of architecture differences, mere porting of such libraries to \textit{mobile devices} yields suboptimal performance or is even impossible in some cases. 
%
%
%
%
%

In this paper, we present PhoneBit, a GPU-accelerated BNN inference engine for Android-based mobile devices that explores both software and hardware-level optimization opportunities of BNNs on mobile GPUs. PhoneBit provides a highly optimized framework for execution of BNNs that explores efficient BNN operators to apply high-level optimizations on mobile devices. Using these operator optimizations, an inference of BNN can be effectively executed with minimal memory footprint during run-time. PhoneBit is implemented as a stand-alone inference engine for BNNs with OpenCL, a GPU programming language supported by most mobile GPU architectures.
To enable real-time and highly efficient BNN implementations on mobile GPUs, we present the detailed parallelization optimization practices when implementing PhoneBit with OpenCL. 
The contributions of this work can be summarized as follows:
\begin{itemize}
\item We present PhoneBit framework for exploiting computing power of BNNs on mobile GPUs in a systematic way.
\item We propose a set of operator-level optimizations including locality-friendly data layout, bit packing with vectorization and layers integration for efficient binary convolution. 
\item We provide a detailed implementation and parallelization optimization for PhoneBit to optimally utilize the memory bandwidth and computing power of mobile GPUs.   
\end{itemize}
We evaluated PhoneBit using real world workloads on two mobile platforms with different SoCs: Snapdragon 820 and Snapdragon 855. Experimental results show that PhoneBit can achieves up to $38\times$ speedups and $89\times$ energy efficiency over existing GPU-based frameworks. 

 
\section{Related Work}
\label{sec:rw}
\noindent{\textbf{Binary Neural Networks.}} 
BNNs are deep neural networks that use binary values for activations and weights, instead of full precision values. BNNs are good candidates for deep learning implementations on FPGAs and ASICs due to their bitwise
efficiency. Accelerating BNN inference at hardware level has been intensively investigated in~\cite{7929192,Nakahara}. However, these solutions are not versatile as they are hardware-specifically dependent. In general, mobile computing requires the versatility on the applications, i.e., applications developed for one vendor’s platform can also execute on other vendors’ platforms. Regarding software-based solutions, BMXNet~\cite{bmxnet} is an open-source BNN library based on MXNet, which is mainly designed for desktop platforms. 
Recently, Pedersoli et al.~\cite{Espresso} presented an BNN inference library Espresso which was written in CUDA. In~\cite{Espresso}, the optimization for binary matrix multiplication kernels was discussed. However, the more advanced optimizations such as layer integrations are not presented yet. In addition, Espresso is \textit{only} tailored for desktop/server GPUs and cannot be applied on resource constrained \textit{mobile devices} because of architecture differences.

\noindent{\textbf{Deep Learning Framework for the Mobile.}} There are many studies designed deep learning frameworks for mobile devices. These frameworks include TensorFlow Lite (TFLite) from Google~\cite{tflite}, Caffe2 from Facebook~\cite{caffe2},  PaddlePaddle Lite from Baidu~\cite{PaddlePaddle}, CoreML from Apple~\cite{CoreML}, and CNNdroid~\cite{Latifi2016CNNdroid}. However, most of the existing mobile libraries are limited to CPU/GPU acceleration for the computations of general CNN models with full or half precision~\cite{ji2019hg}. Currently, TFLite supports 8-bit network quantization \textit{only} for CPUs. Network quantization on GPUs is not supported yet in TFLite. To the best of our knowledge, such GPU-accelerated BNN libraries are not available yet on mobile platforms.

\section{Background}
\label{sec:bg}
\subsection{Mobile GPU Architecture}
\begin{figure}
    \centering
    \vskip -0.3cm
   \includegraphics[clip, trim=8cm 8.2cm 6.5cm 7.5cm, width=0.5\textwidth]{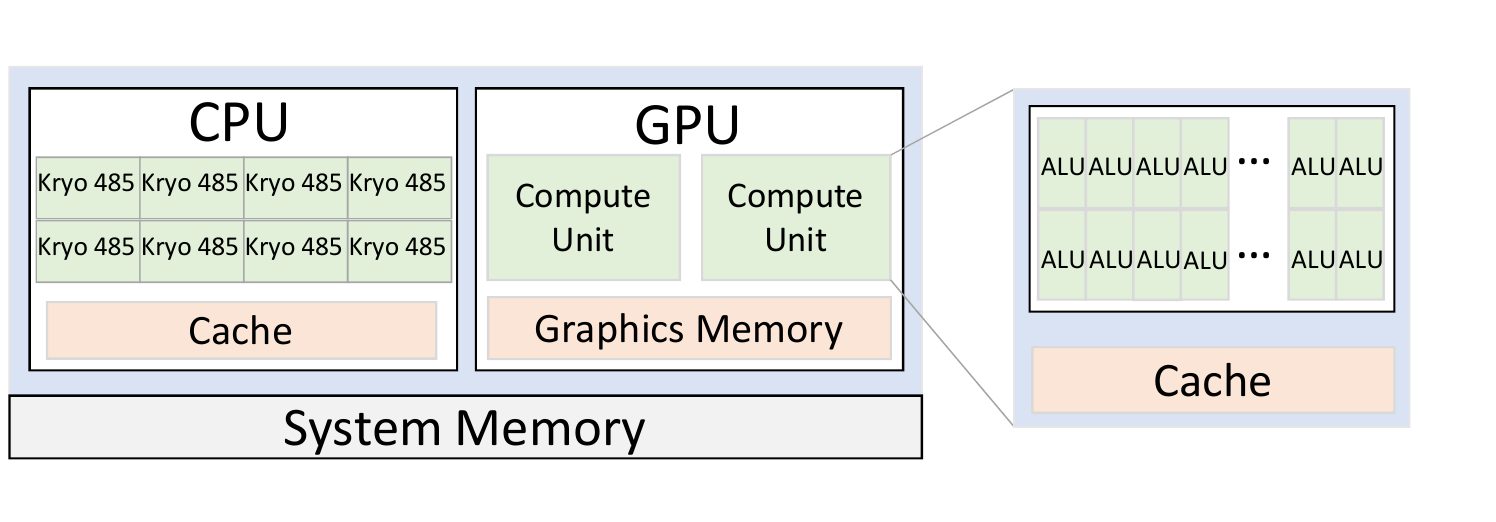}
		\vskip -0.5cm
    \caption{Architecture overview of Qualcomm Snapdragon 855 SoC integrated with Kryo 485 CPU and Adreno 640 GPU.} 
		\vskip -0.5cm
    \label{fig:architecture} 
\end{figure}

Now, emerging programming models such as OpenCL have been supported by mobile GPUs. GP-GPU computing in mobile devices becomes possible. However, due to the strict area and power constraints, mobile GPUs have some big differences from server/desktop GPUs~\cite{Latifi2016CNNdroid}. In mobile GPUs, the ability of powerful parallel computation relies on parallel computing units called compute units (CUs). Each CU is composed of several parallel ALUs. \figref{fig:architecture} illustrates the architecture overview of Qualcomm Snapdragon 855 SoC that integrates a mobile GPU named Adreno 640 consisting of 2 CUs. Each CU in Adreno 640 GPU contains 192 ALUs with SIMD operations.  


%


Compared with server/desktop GPUs, mobile GPUs have limited processing resources. A state-of-the-art desktop GPU usually have thousands of ALUs and thousands of KBytes of L1/shared memory on-chip in total. Meanwhile, a state-of-the-art mobile GPU like Adreno 640 produced by Snapdragon only has 384 ALUs and 1024 KBytes graphics memory on-the chip. Under such limited computational resources, it is more challenging to build high-performance programs on mobile GPUs, especially for DNN-based applications when compared to desktop GPUs.     

%
%
%



\subsection{Binary Convolution Operation}
\label{sec:BCO}
In BNN,  the weights are binarized, which drastically reduces memory size and accesses. To achieve efficient convolution operators, dot production in binary convolution operation can be replaced by $xor$ (for multiplications) and $popcount$ (for accumulations), as presented in \eqnref{eq:xor}. $\vec{A}$ and $\vec{B}$ are binary vectors of length $Len$ while $a_i$ and $b_i$ are the $i_{th}$ binary elements in  $\vec{A}$ and $\vec{B}$, respectively. 
\begin{align}
\label{eq:xor}
\vec{A} \cdot \vec{B} =  Len - 2 \times (popcount(xor(a_i, b_i)))
\end{align}

The input of the convolution layer typically comes as images, which conflicts with the requirement of binary input for the binary convolution layer. In PhoneBit, we follow~\cite{Courbariaux2016Binarized} 
and split the input $I$ into bit-planes $I_i$. Then, binary convolution is operated on bit-plane $I_i$ and binary weights $W$. The output $s$ can be obtained by summing up the convolution of all bit-planes.
\begin{align}
\label{eq:8_gemm}
s = \sum_{n=1}^{8} 2^{n-1}<I_i \cdot  W>
\end{align}
where $I_i$ is the bit-plane after splitting and $<>$ denotes a binary convolution operation. 

\section{The Overview of PhoneBit Framework}
\label{sec:meth}

\begin{figure}
    \centering
    \includegraphics[clip, trim=3.5cm 8cm 3cm 7cm, width=0.49\textwidth]{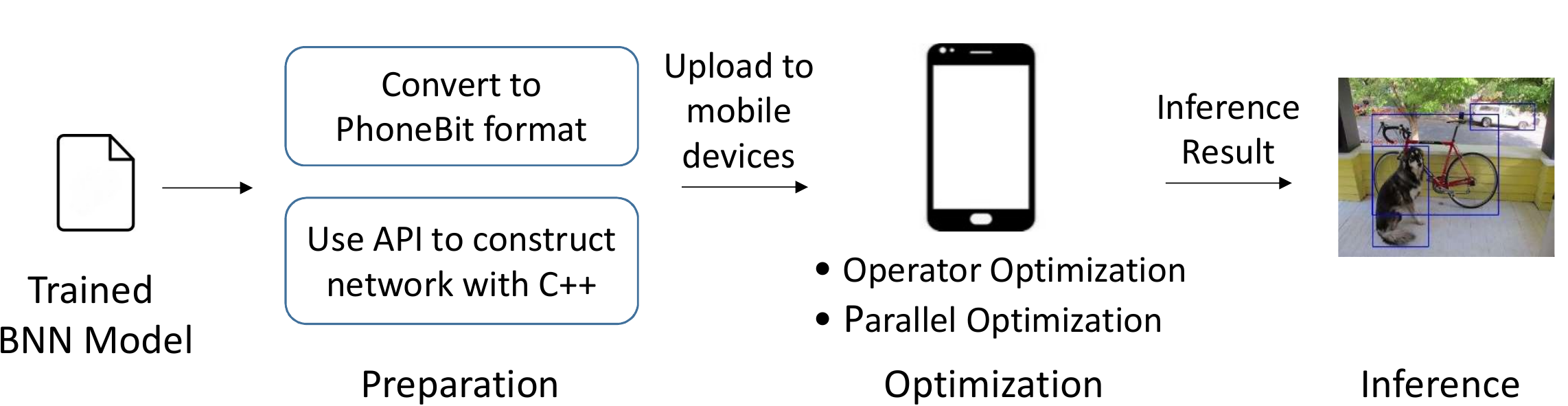}
		\vskip -0.3cm
    \caption{Execution steps in deploying trained BNN models on mobile devices.} 
		\vskip -0.5cm
    \label{fig:overview} 
\end{figure}

This section provides a high level overview of PhoneBit framework that significantly simplifies the deployment of BNN on mobile devices. In the PhoneBit framework, nearly all common types of BNN layers are well supported such as convolution, pooling, batch normalization, and dense (i.e. fully connected) layers. We implement PhoneBit as a stand-alone GPU-accelerated inference engine for BNNs using OpenCL on mobile devices. \figref{fig:overview} summarizes execution steps involved in deploying trained BNN models on mobile devices. PhoneBit first takes a model trained by existing BNN training frameworks and provides a set of scripts to transform it into the compressed PhoneBit format. Once the trained BNN model is generated and uploaded to mobile phones, the BNN model can be deployed on mobile phones in a few simple steps. A code snippet of the corresponding C++ inference interface is presented in \figref{list:code}. 
In a few lines of code, a user can call the PhoneBit APIs to construct networks under the PhoneBit framework and get a deployable module for BNN using GPUs on mobile phones.

\begin{figure}[htp]
\begin{minipage}{1\columnwidth}
\begin{lstlisting}
//(1) Initalization (Preparing Input and Packing Weights)
...
//(2)Construct Network and Forwarding
//layer 1
conv1.bforward_S(&img, &padding_size1, &kernel_size1, &stride1, &w1, &bn1);
pool1.forward_S(&conv1.out, &size1, &stride_p1, MAX);
//layer 2
conv2.bforward64_S(&pool1.out, &padding_size2, &kernel_size2, &stride2, &w2, &bn2);
pool2.forward_S(&conv2.out, &size2, &stride_p2, MAX);
//add more layers
.....
\end{lstlisting}
	\vskip -0.4cm
\caption{Code snippet for the BNN deployment using PhoneBit}
\label{list:code}
\end{minipage}
\end{figure}

\section{Optimizing Binary Operations}
\label{sec:OBO}
PhoneBit framework exploits efficient binary neural network operators to apply high-level optimizations on mobile phones. It implements many operator-level optimizations, including: locality-friendly data layout, which enables efficient memory access; bit packing with vectorization, which packs bits on channel direction; and layer integration, which integrates multiple operations in different layers together. In this section, we will present these optimization techniques for efficient binary neural network operators in detail, with a focus on binary convolution.  
     
\subsection{Channel Compression}
In BNN, matrix multiplications can be efficiently executed by utilizing the binary instructions $xor$ and $popcount$ when working with binarized weights and input data. In PhoneBit, we pack the bits in channel dimensions into multiple compressed bytes (or other data type such as \textit{short} with 16-bit and \textit{int} with 32-bit), and then perform a compressed convolution on compressed tensor and filters.       

\subsubsection{Data Layout} 
In BNN, we require to compress the bits in channel dimensions to achieve compressed tensors for efficient convolution operations. This determines that the channel dimension is the best choice along which bit packing should be conducted. Rather than using the NCHW data layout, the default setting adopted in mainstream deep learning frameworks such as Caffe~\cite{caffe2} and Torch~\cite{Torch7}, PhoneBit use NHWC data layout to achieve efficient bit-packing. In PhoneBit, we use $T=R^{H*W*C}$ to represent the tensor, where $h\in{[0, H]}$, $w\in{[0, W]}$, and $c\in{[0, C]}$ represent the tensor dimensions of height, width, and channel, respectively. A minor-to-major dimension order of tensor $A$ is row-major order with interleaved channels. By using internal data layouts for BNN execution, channel-packing can be effectively performed along the channel dimension with efficient memory access when unrolling a tensor, as well as efficient bit operations when performing convolution computation.
\subsubsection{Bit Packing} Mobile GPU uses a SIMD model with various additional constraints that make it more efficient. OpenCL supports built-in vector data types~\cite{OpenCLSpec} which are defined with the type name, i.e., $uchar$, $ushort$, $uint$, and $ulong$ followed by a literal value $n$ that defines the number of elements in the vector. Supported vectorization values of n are 2, 4, 8, and 16. To maximize vector unit utilization, PhoneBit uses these built-in OpenCL APIs in its kernel code for efficient bit-wise operations. PhoneBit supports parallel bit-wise operations in different parallelization granularity from 8-bit to 1024-bit\footnote{Using ulong16 to achieve 1024-bit vectorization}. To achieve massive parallelism of BNN executions, PhoneBit selects the optimal bit packing strategy and computing kernel according to channel dimensions.  

\subsection{Layer Integration}
\label{sec:LI}
In general, convolutional operations in BNN consist of three layers, including binary convolution, batch-normalization (BN), binarization layers. The layer overflow that transmits data among different layers requires expensive data movement operations that copy and write tensors for several times during the inference. In addition, deploying a BN layer will further introduce floating-point computations and increase the computation burden. To resolve these issues, we propose a layer integration technique that combines multiple operators in these layers into a single kernel without saving the intermediate results in memory. This optimization can greatly reduce execution time for mobile GPUs.

Now, we will mathematically show how these three layers can be aggregated as one integration operator. For ease of presentation, we denote $x_1$ as the result of binary convolutional computation without considering the bias $b$ and $x_2$ as the output of binary convolution layer. Therefore, we have:  
\begin{align}
\label{eq:x2}
x_2=x_1+b 
\end{align}
In the BN layer, let $x_3$ be the output of a BN layer, $\gamma$ and $\beta$ denote the trainable parameters, $\mu$ and $\sigma$ are estimated from the sample mean and sample variance of mini-batch, respectively. The transformation can be presented as follows:
\begin{align}
\label{eq:x3}
x_3=\gamma\cdot{}\frac{x_2-\mu}{\sigma}+\beta 
\end{align} 
Substituting $x_2$ in \eqnref{eq:x3} using \eqnref{eq:x2}. $x_3$ can be represented as follows:
\begin{align}
\label{eq:3_1}
x_3=\frac{\gamma}{\sigma}\cdot{(x_1 - \xi)}
\end{align} 
where $\xi$ is determined as\footnote{According to \cite{networkslimming}, the convolutional channel with $\gamma$ = 0 can be pruned. Thferefore, we do not consider the case with $\gamma$ = 0. }:
\begin{align}
\label{eq:xi}
\xi=\mu-\frac{\beta\cdot{\sigma}}{\gamma}-b
\end{align} 
Note that $\xi$ can be computed in the off-line stage without increasing the runtime computation burden. Then, $x_3$ is binarized to the output $x_4$ after a BN layer. Therefore, we have: 
\begin{align}
\label{eq:sign}
x_4 &=
\begin{cases}
1        & \text{if } x_3 \geq 0 \\
0        & \text{otherwise}
\end{cases}
\end{align}
From \eqnref{eq:3_1}, we know the parameters of $\gamma$ and $\xi$ determine the sign of $x_3$. Therefore, we can easily transform the transformation functions above as an integrated one, as presented in \eqnref{eq:bn0}.


\begin{align}
\label{eq:bn0}
x_4 &=
\begin{cases}
1        & \text{if } x_1 \geq \text{$\xi$ and }\gamma > 0\\
0        & \text{if } x_1 < \text{$\xi$ and }\gamma > 0 \\
1        & \text{if } x_1 \leq \text{$\xi$ and }\gamma  < \text{0}\\
0        & \text{if } x_1 > \text{$\xi$ and }\gamma < \text{0}
\end{cases}
\end{align} 




\section{PhoneBit Implementation and Parallelization Optimization}
\label{sec:PO}
In this section, we provide a highly optimized GPU-accelerated implementation for PhoneBit. We implement PhoneBit as a stand-alone inference engine for BNNs with OpenCL, a GPU programming language supported by most mobile GPU architectures. Major optimization steps, such as workload optimization, avoiding branch divergence, memory optimization, are provided to demonstrate the effectiveness of the optimization practices when
implementing PhoneBit. 
\subsection{Memory Optimization}

Optimization techniques for the computational efficiency of the binary operations have been discussed in \secref{sec:OBO}. PhoneBit has intricate data structures and their memory behavior has significant impact on the performance. In this section, we focus on the memory efficiency of PhoneBit and present the following memory optimization techniques used in PhoneBit.  
\subsubsection{Vectorized Load/Store}
We first use vectorized load/store functions in OpenCL to greatly reduce the number of load/store operations in PhoneBit. These build-in functions in OpenCL, supported by most mobile GPUs, allow the hardware to bulk load/store data to/from memory. Such strategy typically takes advantage of the spatial and temporal locality properties of the programs. To achieve better bandwidth utilization, PhoneBit loads/stores the data in chunks of multiple bytes (e.g., 128-bit) using vectorized load/store functions~\cite{Wang:2018:OOB:3204919.3204935}.  

\subsubsection{Coalesced Memory Access}
To maximize memory access bandwidth, the GPUs try to coalesce memory accesses from work items in the same wavefront into a single memory request if these memory accesses have good spatial localities. To achieve the optimized performance, 
we apply the adjacent memory-based tile optimization~\cite{Wang:2018:OOB:3204919.3204935} to achieve the coalesced memory access.
In addition, for binary convolution operations on the NHWC data layout, the binary operations in each region of the feature map are directly applied to the packed bits that are stored in memory consecutively. Therefore, memory coalesced accesses can also be achieved along the lowest channel dimension in PhoneBit. 

\subsubsection{Memory Latency Hiding}
Latency hiding refers to the process of overlapping memory operations with computation to maximize the utilization of memory and compute resources. Latency hiding is one of the most powerful characteristics of GPU for efficient parallel processing and enables GPU to achieve high throughput. In PhoneBit, we interleave the memory load and computation into different threads such that memory load and computation can be swapped in a pipelined manner to achieve characteristic of the latency hiding. The pipeline can hide most memory access overheads and almost fully utilize compute resources.    

\subsection{Workload Optimization}
In BNN, we pack the binary output of convolution kernels to achieve the compressed feature maps. One straightforward approach is to use one thread to conduct the calculation of one convolution kernel and use a byte to store the binarized output. Then, one additional thread is applied to pack the bits along the channel dimensions. However, such implementation requires several individual threads to perform channel compression procedures. The computations of convolution, BN, binarization layers cannot be aggregated for efficient binary operators.      
Furthermore, this implementation also introduces additional cost for thread synchronization.

\begin{figure}
    \centering
    \vskip -0.1cm
    \includegraphics[clip, trim=2.5cm 8cm 8cm 4cm, width=0.49\textwidth]{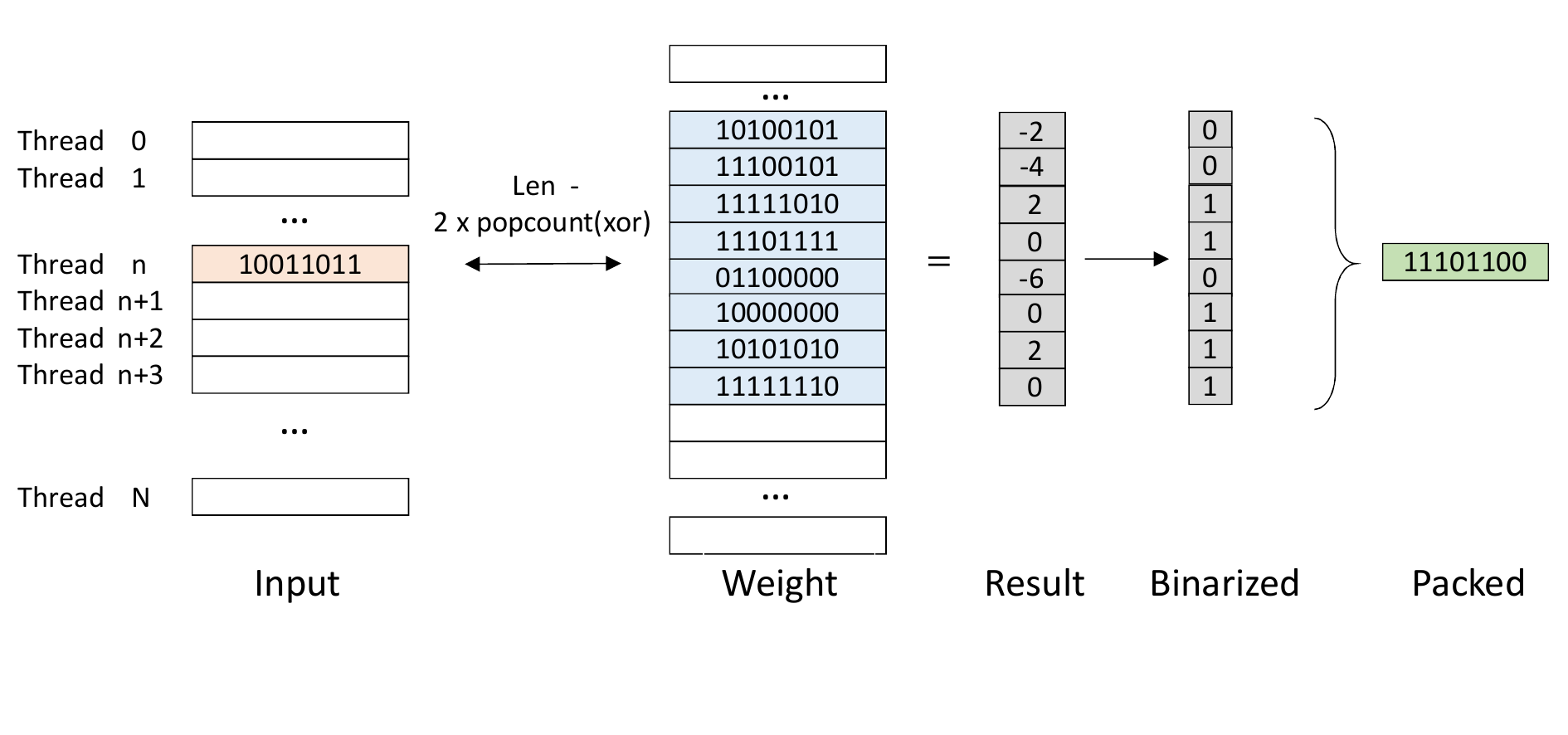}
		\vskip -0.9cm
    \caption{One thread computes 8 filters, binarizes 8 results and pack into one byte. $Len$ is the length of vector.} 
		\vskip -0.6cm
    \label{fig:workload} 
\end{figure}

To resolve the issues above, PhoneBit uses a single thread to calculate the binary output of 8 convolution kernels and pack them in a compressed byte, as shown in \figref{fig:workload}. In this strategy, the data are stored in private memory of the thread and can be rapidly accessed. Therefore, packing operations can be integrated to avoid unnecessary memory access operations as well as thread synchronization cost. However, due to the limitation of private memory size, one thread cannot load too much data once a time. Especially when the channel number is too large, private memory of one thread cannot load the required data for the computation of 8 convolution kernels. In PhoneBit, we assign the computing thread with proper workload according to the number of channels. When the channel number is no greater than 256, PhoneBit will conduct the computations for 8 convolution kernels and integrate packing operations inside. Otherwise, packing operations will be processed separately without integration.

\subsection{Avoiding Branch Divergence}
Generally, GPUs are not efficient when work items in the same wave follow different execution paths. For divergent branches, some work items may have to be masked, resulting in lower GPU occupancy~\cite{Wang:2018:OOB:3204919.3204935}. In PhoneBit, the computations of BNN layers have been transformed as \eqnref{eq:bn0}, which requires to use divergent checks to determine the output. In PhoneBit, we present novel software-based optimizations to convert divergent check operations to fast logic operations. We first use the truth table to represent the determination of \eqnref{eq:bn0}. Based on this representation, the following logic function \eqnref{eq:bn2} can be computed by the laws of boolean algebra and be further simplified by Karnaugh maps~\cite{Karnaugh1953}.

\begin{align}
\label{eq:bn2}
x_4 = (A\ xor\ B)\ or\ C
\end{align} 
where $A$, $B$, and $C$ denote the boolean variables of $x_1< \xi$, $\gamma > 0$, $x_1= \xi$, respectively, which can be rapidly determined using OpenCL build-in functions $isless$, $isgreater$ and $isequal$. By using fast logic operations in \eqnref{eq:bn2}, divergent checks in PhoneBit can be avoided.

\section{Experiment}
\begin{table}
\centering
\caption{Mobile devices}
\vskip -0.2cm
\label{ex:rw}
{   \scriptsize
    \resizebox{0.5\textwidth}{!}{
        \begin{tabular}{c|c|c|c|c|c}
        \hline
        Device   &  SOC           & Memory     & OS            & OpenCL Version    & ALUs in GPU   \\ 
        \hline
        Xiaomi 5 & Snapdragon 820 & 3GB             & Android 7.0   &  2.0              &    256       \\   
        Xiaomi 9 & Snapdragon 855 & 8GB             & Android 9.0   &  2.0              &    384       \\  
        \hline
        \end{tabular}
    }
}
\vskip -0.4cm
\end{table}

\noindent{\textbf{Experiment Setup.}} PhoneBit is evaluated on two mobile devices with different SoCs: Snapdragon 820 and Snapdragon 855. The detailed hardware configurations are shown in \tabref{ex:rw}. We evaluate PhoneBit with three classic neural network models, namely, AlexNet network for CIFAR10 dataset, YOLOv2 Tiny network for VOC2007 dataset and VGG16 network for CIFAR10 dataset. We conduct the experiments to evaluate the efficiency of PhoneBit by the extensive comparison with state-of-the-art frameworks:
\begin{itemize}
\item \textbf{CNNdroid~\cite{Latifi2016CNNdroid}:} a RenderScript-based framework~\cite{RenderScript} for paralleling computations of full-precision CNN across CPUs and GPUs. Both CPU- and GPU-based executions are evaluated for performance comparison. However, as indicated in~\cite{Benchmark}, even executing in GPU-based execution, RenderScript is not always using GPUs on all the devices \text{--} sometimes it is still running on a CPU only. 
\item \textbf{TensorFlow Lite (TFLite)~\cite{tflite}:} a lightweighted deep learning framework developed by Google for mobile devices. TFLite supports accelerators on both CPUs and GPUs. Currently, TFLite only support lower to 8-bit quantization specification on CPUs and does not support model quantization on GPUs~\cite{tflite}. Due to these limitations, three implementations, including one GPU-based and two CPU-based executions with and without quantization, are evaluated for performance comparison. 
\end{itemize} 
We implement forwarding of the benchmark networks on the mobile devices. The metrics of
accuracy, model size, runtime, power and energy consumption are evaluated and compared to demonstrate the efficiency of PhoneBit.  

\begin{table}
    \centering
    \caption{Model size(MB) and precisions}
		\vskip -0.2cm
    \label{tab:Table_model_size}
    {   \scriptsize
    \resizebox{0.5\textwidth}{!}{
    \begin{tabular}{c|c|c|c|c}
    \hline
    \multirow{2}{*}{} &\multicolumn{2}{c|}{\multirow{2}{*}{Model Size(MB)}}&\multicolumn{2}{c}{\multirow{2}{*}{Precision(\%)}}\\
                       &\multicolumn{2}{c|}{} &\multicolumn{2}{c}{} \\
    \hline
                    &Full-precision CNN  &BNN  &Full-precision CNN & BNN  \\ 
    \hline

    AlexNet             &249.5            &16.3        &89.0      & 87.2  \\
    YOLOv2 Tiny         &63.4             &2.4        &57.1      & 51.7 \\
    VGG16               &553.4            &32.1       &92.5      & 87.8  \\          
    \hline
    \end{tabular}}
    }
    \vskip -0.4cm
\end{table}

\noindent{\textbf{Accuracy and Model Size.}} 
First, we run three binarized benchmark networks using PhoneBit, and compare their accuracy and model size with full-precision networks. The compared results are shown in \tabref{tab:Table_model_size}. The model size of networks is critical for space-constrained mobile devices. For all three benchmark networks, PhoneBit can achieve significant compression rate with an acceptable accuracy loss.The model size of PhoneBit is on average $19.6\times$ smaller than full-precision networks. For large network model such as VGG16, PhoneBit can achive $17.2\times$ storage efficiency with $4.7\%$ accuracy loss when compared to full-precision networks.   

\begin{table*}  
    \scriptsize
    \caption{Average runtime(ms) in Snapdragon 820 and Snapdragon 855 platforms using different frameworks}
		\vskip -0.3cm
    \resizebox{\textwidth}{!}{
    \label{tab:Table_time_cost}
    \begin{tabular}{c|c|c|c|c|c|c|c|c|c|c|c|c}
    \hline
    \multirow{4}{*}{}   &\multicolumn{6}{c|}{\multirow{2}{*}{Snapdragon 820}}   & \multicolumn{6}{c}{\multirow{2}{*}{Snapdragon 855}}  \\     
                        & \multicolumn{6}{c|}{}                                 & \multicolumn{6}{c}{} \\\cline{2-13}
                        & \multicolumn{2}{c|}{CNNdroid} & \multicolumn{3}{c|}{Tensorflow Lite} & \multirow{2}{*}{PhoneBit} & \multicolumn{2}{c|}{CNNdroid} & \multicolumn{3}{c|}{Tensorflow Lite} & \multirow{2}{*}{PhoneBit} \\\cline{2-6} \cline{8-12}
                        & CPU & GPU &CPU & GPU & CPU Quant &  & CPU & GPU &CPU & GPU & CPU Quant & \\
    \hline
    AlexNet             & 8243 & 766 & 143 & CRASH & 103 & 22.9     & 5621 & 369 & 87 & CRASH & 24 & 9.8        \\
    YOLOv2 Tiny         & 51313 & 1483 & 669 & 468 & 503 & 42.1     & 23144 & 845 & 306 & 430 & 88 & 22.6       \\
    VGG16               & OOM & OOM & 2607 & CRASH & 1907 & 152.3   & OOM & OOM & 932 & CRASH & 252 & 73.8       \\
    \hline
    \end{tabular}
    }
		 \vskip -0.5cm
    
    
\end{table*}

\noindent{\textbf{Runtime Performance.}} 
Next, we will demonstrate the efficiency of PhoneBit by comparing runtime performance with state-of-the-art frameworks. \tabref{tab:Table_time_cost} shows the runtime performance of three benchmark networks for the compared frameworks under different hardware settings. From \tabref{tab:Table_time_cost}, we can make the following observations: (1) PhoneBit brings significant speedups over all of the compared frameworks on both mobile devices. Compared with CNNdroid~\cite{Latifi2016CNNdroid}, PhoneBit on average gains $794\times$ and $35\times$ speedups on CPU based and GPU based executions, respectively. On TFLite, PhoneBit can on average bring $12\times$, $15\times$, and $6\times$ faster speedup over CPU based, GPU based and CPU quantization based executions, respectively. (2) Stability issues have been observed on CNNdroid and TFLite frameworks. 
For large-scale network VGG16, CNNdroid runs out of memory (OOM) on both CPU and GPU implementations while TFLite runs into crash (CRASH) and fails to produce the results on GPU based executions. In contrast, PhoneBit can work flawlessly with all benchmark networks with faster inference speed. 
%
%
%
%
%
%
%
%
%
%

\begin{figure}
    \centering
    \vskip -0.1cm
    \includegraphics[width=0.5\textwidth,height=0.3\textwidth,trim=0 5 0 30,clip]{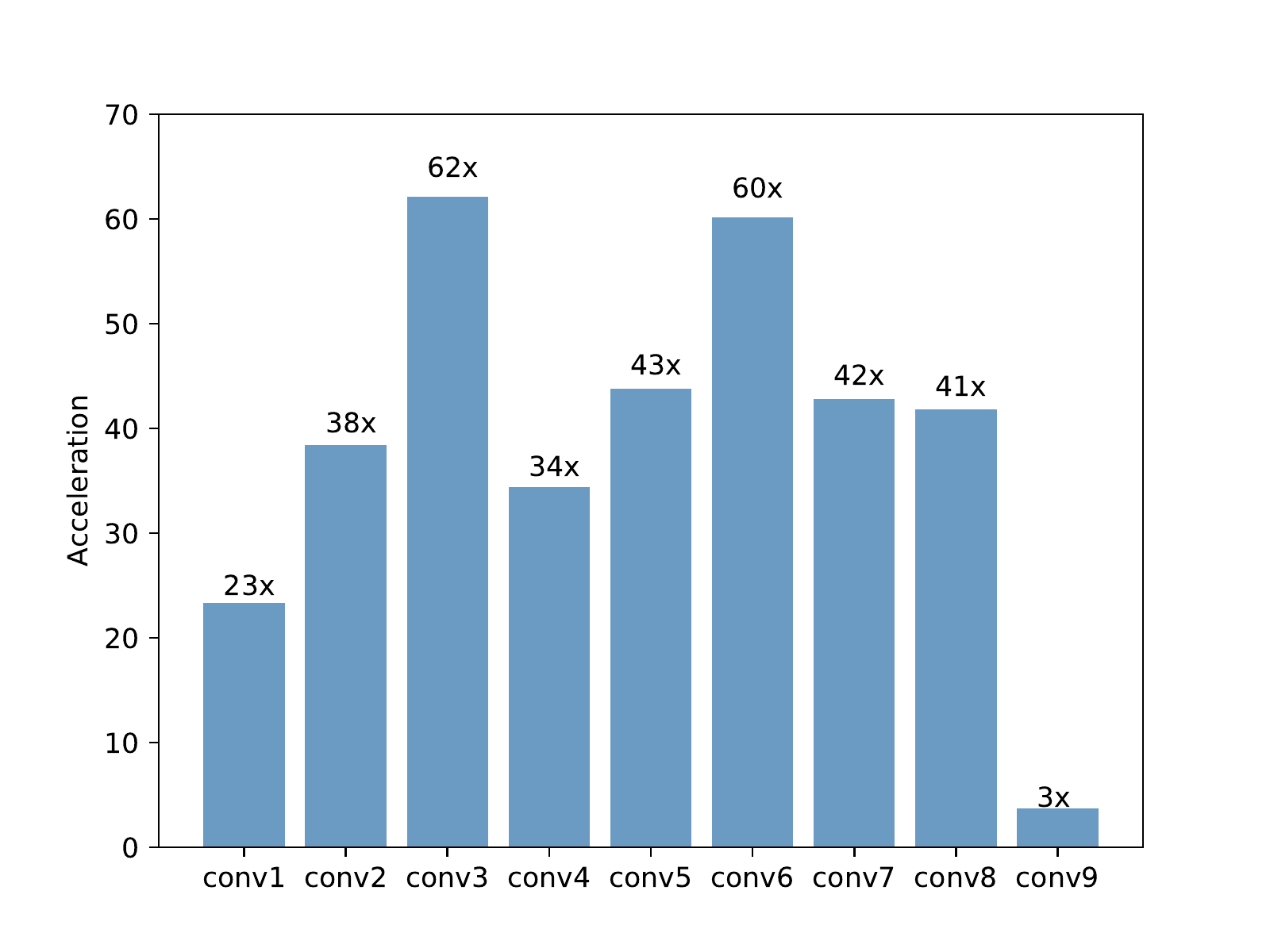} 
		 \vskip -0.6cm
    \caption{Performance improvement brought by PhoneBit BNN implementations, with counterpart float-value operators in CNNdroid (GPU based execution) as the baseline(1x), tested on Snapdragon 855 platform.} 
    \label{fig:Figure_Acceleration_In_Conv} 
		 \vskip -0.6cm
\end{figure}

We also investigate how our parallelism optimization methods can explore the massive parallelism on various convolutional layers of networks. To examine the impact of PhoneBit on various layers of networks, we measure run-time of each layer in YOLOv2-Tiny where the first layer comes as 8-bit integers and the last layer is a full precision layer for final float type output.  
In PhoneBit, multiple operators in binary convolutional, BN, binarization layers are combined into a single kernel for efficient convolutional computation, using layer integration technique presented in \secref{sec:LI}. \figref{fig:Figure_Acceleration_In_Conv} shows the performance improvement over CNNdroid with GPU based execution\footnote{TFLite provides a static compiled library in Android and therefore cannot measure run time in single layer} on the integrated layers for YOLOv2-Tiny network. For the middle binary layers from conv2 to conv8, PhoneBit can bring $45\times$ (up to $62\times$) acceleration over CNNdroid on average. Such acceleration benefits from not only operator optimization techniques presented in \secref{sec:OBO} but also parallelization optimization techniques presented in \secref{sec:PO}. On conv1, PhoneBit only gains about $23\times$ speedups because of additional process of splitting input integer into bit-planes as we described in \secref{sec:BCO}. On the last layer conv9 with full precision, PhoneBit still gains about $3\times$ acceleration over CNNdroid due to the fact of using SIMD operation on build-in dot product function $dot$ in OpenCL.

\noindent{\textbf{Energy Efficiency.}} Energy efficiency is a major design concern for energy-restricted mobile devices. Now, we demonstrate the energy efficiency of PhoneBit by reporting power consumption and energy efficiency under different frameworks. The Trepn Power Profiler~~\cite{qualcomm_trepn_profile} is an on-target power and performance profiling tool for Android mobile devices. We use Trepn Power Profiler to measure the power consumption on Snapdragon 820 platform\footnote{Trepn Profiler on Snapdragon 855 platform cannot obtain the profiling data of battery power.}. \tabref{tab:Table_Energy} shows the power consumption and energy efficiency for YOLOv2 Tiny network. From \tabref{tab:Table_Energy}, we can see that PhoneBit can achieve significant power efficiency when executing YOLOv2 Tiny network. PhoneBit consumes around $226mW$ power and obtains $105.26$ FPS per watt. Compared with CNNdroid and TFLite implementations, PhoneBit brings $4\times$\text{--}$ 1218\times$ faster execution speed, $2\times$\text{--} $4\times$ lower power dissipation and $24\times$\text{--}$ 5263\times$ better performance per power efficiency.

\begin{table}
    \centering
    \caption{Energy consumption per image frame for YOLOv2 Tiny network on Snapdragon 820 platform}
		\vskip -0.2cm
    \label{tab:Table_Energy}{   
        \scriptsize
        \resizebox{0.5\textwidth}{!}{
            \begin{tabular}{c|c|c|c|c|c|c}
            \hline
            \multirow{2}{*}{} & \multicolumn{2}{c|}{CNNdroid} & \multicolumn{3}{c|}{Tensorflow Lite} & \multirow{2}{*}{PhoneBit} \\ \cline{2-6}
             
                            &CPU    & GPU &CPU & GPU & CPU Quant &\\\hline
            
            Watts(mW)       & 914 & 573 &626 & 540 & 452 & 225.67\\
            Efficiency(FPS/W)  &0.02 &1.18 &2.39 & 3.97& 4.40&105.26 \\
						\hline
            \end{tabular}
        }
    }
    \vskip -0.6cm
\end{table}
\vskip -0.3cm

\section{Conclusion}
\label{sec:conclusion}
In this paper, we propose a GPU-accelerated BNN inference engine PhoneBit for Android-based mobile devices. In PhoneBit, we propose a set of operator-level optimization techniques to fully explore computing power of BNNs on mobile GPUs. To enable real-time and highly efficient BNN implementations on mobile GPUs, we present the detailed parallelization optimization practices when implementing PhoneBit with OpenCL. On the evaluations of popular neural networks, PhoneBit achieves significant speedups and energy efficiency when compared with state-of-the-art frameworks for mobile devices.

{
\footnotesize
\bibliographystyle{abbrv}
\bibliography{../bib/mc_1}
}

\end{document}